\documentclass[11pt]{article}
\usepackage{moriond,epsfig}

\def\be{\begin{equation}}
\def\ee{\end{equation}}
\def\bea{\begin{eqnarray}}
\def\eea{\end{eqnarray}}
\begin{document}
\vspace*{2cm}
\title{THE HEAVY-ION PROGRAMME OF THE ALICE EXPERIMENT AT LHC}

\author{ A. MASTROSERIO on behalf of the ALICE Collaboration}

\address{Universit\'a Degli Studi Di Bari and INFN (Bari)\\
 Via E. Orabona 4, 70126 Bari, Italy}

\maketitle \vspace*{1cm} \abstracts{The ALICE experiment at LHC is
mainly dedicated to heavy-ion physics. An overview of its
performances, some predictions related to its first measurements and
QGP observable measurements will be given.} \vspace*{0.5cm}
\section{Introduction}
The heavy-ion programme of the ALICE experiment aims to study
nuclear matter at extreme conditions as very high energy densities
and temperatures. Lattice QCD calculations predict that above a
critical temperature $T_c~\sim$ 170 MeV and a critical energy
density $\epsilon_c~\sim~1$ GeV/fm$^3$, the nuclear matter undergoes
a phase transition to the Quark Gluon Plasma (QGP) state
\cite{pprI}. The heavy-ion programme at LHC will start with Pb-Pb
collisions at $\sqrt{s_{NN}}=~5.5$ TeV where energy densities well
above the critical value $\epsilon_c$ will be reached. Besides the
phase transition, other phenomena like hadronic collective flow and
new mechanisms at the hadronization stage (e.g.: quark coalescence)
are also expected to occur. The ALICE experiment will investigate
the features of strongly interacting matter in a new energy domain,
hopely leading to unquestionable signatures of the QGP state. Its
physics programme will cover many topics in different physics
domains to cope with the study of a very complex system with many
degrees of freedom. An important benchmark will be provided by pp
collisions and also by smaller collision systems as pPb, dPb,
$\alpha$Pb and lighter nuclei collisions (e.g.: Ar) \cite{pprI}. \\
In the next sections, an overview of the ALICE detector capabilities
and predictions within first heavy-ion measurements will be given.
Also some interesting QGP driven observables that can be measured in
ALICE will be shown.

\section{The ALICE experiment at the LHC}
The ALICE set up consists of a central part and a forward part on
one side for muon detection (Muon Arm). The former is inside a
magnet, the L3 magnet, which can provide a solenoidal magnetic field
between 0.2 T and 0.5 T.  Smaller detectors are also present at very
forward angles to measure global event characteristics. The tracking
detectors, such as the Time Projection Chamber (TPC), the Inner
Tracking System (ITS), the Time Of Flight detector (TOF) and the
Transition Radiation Detector (TRD), are located in the central part
and they have full azimuthal coverage and a pseudorapidity coverage
$|\eta| < 0.9$. Inside the magnet there are also smaller acceptance
detectors such as the High Momentum Particle Identification (HMPID,
based on the Cherenkov detectors), which identifies charged hadrons
between 1 - 5 GeV/c, a single arm electromagnetic calorimeter (PHOS)
and a Photon Multiplicity Detector (PMD). Recently a new
electromagnetic calorimeter (EMCAL), which will be located inside
the L3 magnet, has been included in the ALICE experiment.\\
The detectors have been designed to provide tracking and particle
identification capabilities to cope with a primary charged particle
multiplicity in central Pb-Pb collisions up to dN/dy = 8000 at
midrapidity. Studies performed at dN/dy=6000 have shown a very good
momentum resolution, less than 0.8 \% at $p_T <$ 2 GeV/c up to 3 \%
at $p_{T} \sim$ 100 GeV/c, with a resolution, on the distance of
closest approach of the track to the interaction vertex (DCA
resolution), which is better than 60 $\mu m$ at $p_T >$1 GeV/c. The
ALICE charged particle identification capability allows to identify
pions, kaons and protons on a track-by-track basis between 0.15-5
GeV/c (also up to 50 GeV/c in the relativistic rise) and electrons
well above 1 GeV/c. Also strange particle detection (e.g.:
$\Lambda$) and resonance identification (e.g.: $\rho(770)$ and
$\phi(1020)$ mesons) have been studied in detail and an important
result is that the measured resonance mass resolutions stays below
few MeV/c$^{2}$.

\section{Early measurements: multiplicity and hadron ratios}

The very first measurement envisaged by the ALICE heavy-ion physics
programme will be the charged particle multiplicity at midrapidity
followed by its behaviour along the pseudorapidity range covered by
ALICE. So far, different expectations at LHC energies have been
predicted based either on models or on extrapolations of current
measurements \cite{pprI}. Nowadays, the expected value at
midrapidity ranges between 1200 \cite{eskola} and 2900
\cite{lastcall}. According to the Bjorken scenario\cite{Bjorken}, it
is interesting to infer an estimate of the energy density within the
nuclei overlapping region which should be reached at LHC energies.
At the Relativistic Heavy ion Collider (RHIC), where Au Au
collisions were delivered at $\sqrt{s_{NN}}=200$ GeV), such energy
density was measured as $\epsilon$ = 15 GeV/fm$^3$ ~\cite{phenix}.
Using the same Bjorken formula and assuming a formation time for the
two incoming lead nuclei of the order of 0.2 fm/c, the energy
density values at LHC are 3-5
bigger than at RHIC. \\
Another early measurement of ALICE will be the identified hadron
relative abundances. A very successful model, the thermal model
\cite{equil}, shows that hadron ratios follow a statistical pattern
within a large $\sqrt{s_{NN}}$ interval (2-200 GeV). The interacting
system is considered as a grand canonical ensemble of hadron
resonance gas and particles are formed at a \textit{chemical freeze
out} stage. The model depends on the temperature and the
bariochemical potential and it predicts at LHC energies a chemical
freeze out at the temperature $T_{ch} = 161 \pm 4$ MeV and at a
bariochemical potential $\mu_{ch}=0.8^{+1.2}_{0.6}$ MeV
\cite{lastcall}. Despite the excellent description of the thermal
model of hadron abundances, the dynamics which leads to the
equilibrium is still not clear. A different approach to estimate
hadron abundances is based on the assumption that the expected
increase of strangeness in ultrarelativistic heavy-ion collisions
could deviate from the grand canonical description
\cite{motivstrange}. Another model, the statistical hadronization
model \cite{shm}, has introduced the strangeness phase space
occupancy $\gamma_s \neq 1$ (which implies a deviation from the
equilibrium condition in the strange sector) to describe particle
ratios. The model assumes a sudden hadronization with a $\gamma_s >
1$ which implies a super cooling effect on the system. Further
studies show that if 3 $< \gamma_s <$ 5 the temperature at the
freeze out varies respectively in the range 135 $<T_c<$ 125
\cite{share}. Both multiplicity and relative hadron abundances will
provide the first constraints on the models developed so far and
also new insights on the mechanisms for particle production in
ultrarelativistic heavy-ion collisions.

\section{R$_{AA}$: hadron sector and jet sector}

The study of the inclusive hadron production is very important to
probe the QGP via the parton energy loss dynamics in the medium. In
particular, the nuclear modification factor $R^{h}_{AA}$
characterizes medium induced effects. If h is an hadron specie,
$R_{AA}^{h}$ is defined as: \be R^{h}_{AA}(p_{T},\eta,centrality)
 =\frac{\frac{dN^{AA\rightarrow}h}{dp_Td\eta}}{<N_{coll}^{AA}>\frac{dN^{pp\rightarrow h}}{dp_{T}d\eta}}
\ee At a given collision centrality, it quantifies the ratio between
the measured yield of the hadron h in AA collisions and its yield in
pp collisions, scaled by the factor $N^{AA}_{coll}$ which is the
number of binary nucleon nucleon collisions at that specific
centrality given by Glauber-type estimates. The formation of a new
QCD medium implies that the produced hadrons come from hard
scattered partons which traverse the medium before fragmenting in
the vacuum. The deviation of the $R_{AA}$ from unity is a signal of
different dynamics of parton propagation in heavy-ion collision with
respect to pp collisions. Its suppression factor as high as five in
most central AuAu collisions at RHIC has indicated the formation of
a dense medium where partons lose more energy than in normal nuclear
matter. The medium induced energy loss is related to a medium
dependent parameter, \^q ([\^{q}]= GeV$^2$/fm), that represents the
average squared transverse momentum transferred to a hard parton per
unit path length during its multiple scatterings within the medium
\cite{pprII}. At LHC the kinematical accessible range will be wider
than at RHIC and for light hadrons it is expected $R_{AA}$ = 0.1 at
$p_{T} <$ 20 GeV/c followed by a slow rise up to 0.4 at $p_T <$ 400
GeV/c \cite{lastcall}$^{,}$ \cite{wiedemann} due to the quark
contribution. The color factor, infact, makes gluon loosing energy
more rapidly than quarks.\\
\begin{figure}
 \centering
\includegraphics[height = 6 cm, width = 10 cm]{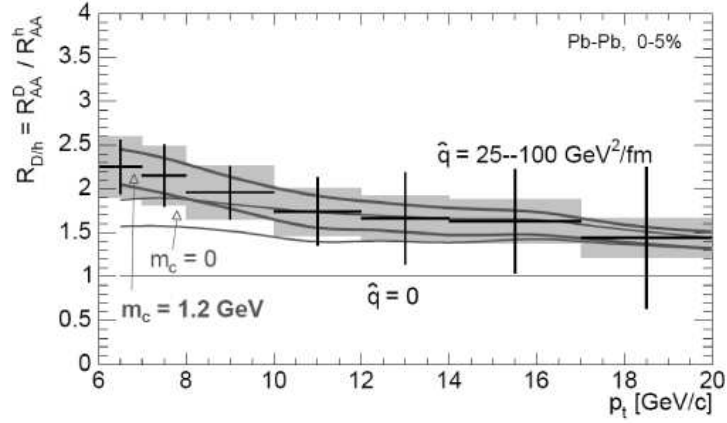}
  \caption{Heavy to light ratio for D$^0$ mesons and charged hadrons in most central Pb-Pb collisions within a \^{q} interval 25-100 GeV/fm$^{3}$.
  The lines define the measured band corresponding to different charm masses. The error bars are the statistical errors,
  the shaded bands corresponds to the systematic errors. Both are related to the $m_c=$ 1.2 GeV case.}
  \label{fig1}
  \vspace*{1cm}
\end{figure}
Furthemore among quarks, the heavy quark gluon radiation in the
medium is different from the others, in particular it should be
smaller than the light quarks as a consequence of a mechanism known
as the dead cone effect \cite{deadcone}. At the same parton energy
the higher is the quark mass, the less is the radiated energy.
Within this scenario, an interesting measurement aiming to probe
parton dynamics in the medium is the heavy to light ratio,
$R_{AA}^{D/h}$ = $R_{AA}^D/R_{AA}^h$, which probes the \textit{color
charge} dependent energy loss. Figure \ref{fig1} \cite{pprII} shows
the ALICE sensitivity to this measurement at different values of
\^{q} and charm quark mass. Another interesting measurement is the
$R_{AA}^{B/D}$ which can probe the \textit{quark mass} dependence of
the partonic energy loss.\\
The comparison of heavy-ion yields with respect to pp yields can be
extended to the jet physics domain. Due to the energy loss in the
medium, in fact, in heavy-ion collision the initial energy of fast
partons is degraded in the medium with a subsequent broadening of
the parton shower along the fast parton direction and a modified jet
structure: the soft particle yield should increase whereas its high
$p_T$ particle yield would decrease \cite{jetmodif}. In ALICE, event
by event jet reconstruction is feasible and to quantify such an
effect an observable which is under study is $R_{AA}(\xi)$, defined
as: \be R_{AA}(\xi)=\frac{1/N^{AA}_{jet}
dN^{AA}/d\xi}{1/N^{pp}_{jet} dN^{pp}/d\xi} \ee where $\xi =
ln(E_{jet}/p^{hadron})$ and $E_{jet}$ and $p^{hadron}$ are the total
jet energy and the jet hadron momentum respectively. This value
would differ from unity in case of medium formation. Ongoing studies
on reconstructed jets in ALICE (also with the introduction of the
EMCAL) show the expected reduction of $R_{AA}(\xi)$ at low $\xi$
(fast hadrons) and its increase at higher $\xi$ where the influence
of systematic errors due to the background source is relevant.

\section{Conclusions}

The ALICE heavy-ion program at LHC energies will be devoted to
understand hadronic matter at extreme conditions and the formation
of the QGP state. Only a few topics have been discussed as first
measurement expectation values and some interesting observables
strictly related to the partonic energy loss in a deconfined medium
as the nuclear modification factor $R_{AA}$ for light flavoured
hadrons, the heavy to light ratio $R_{AA}^{D/h}$ and its extension
to the jet physics domain with the observable $R_{AA}
(\xi)$.\\
The ALICE configuration at the LHC start up in Summer 2008 will
consist of fully operating ITS, TPC, TOF, HMPID, Muon Arm, and
triggering detectors and partially installed TRD, PHOS, PMD
detectors.


\section*{References}
\begin{thebibliography}{99}
\bibitem{pprI} ALICE: Physics Performance Report, Volume I. J. Phys. G: Nucl. Part. Phys. 30 (2004) 1517–1763
\bibitem{pprII} ALICE: Physics Performance Report, Volume II. J. Phys. G: Nucl. Part. Phys. 32 (2006) 1295–2040
\bibitem{eskola} K. J. Eskola, H. Honkanen, H. Niemi, P. V. Ruuskanen, S. S. Rasanen. Phys. Rev. C 72: 044904, 2005
\bibitem {lastcall} N. Armesto et al. J. Phys. G 35 : 054001, 2008
\bibitem{Bjorken} J. D. Bjorken, Phys. Rev. D 27 1983 (140)
\bibitem{phenix} PHENIX Collaboration (K. Adcox et al.). Nucl. Phys. A 757 : 184-283, 2005
\bibitem{deadcone} Yuri L. Dokshitzer, D. E. Kharzeev. Phys. Lett. B 519 : 199-206, 2001
\bibitem{equil} A. Andronic, P. Braun-Munzinger, J. Stachel. Nucl. Phys. A 772: 167-199, 2006
\bibitem{motivstrange} J. Letessier, J.Rafelski. Phys. Rev. C 73: 014902, 2006
\bibitem{shm} J. Letessier, J. Rafelski. Eur. Phys. J. A 35: 221-242, 2008
\bibitem{share} B. Hippolyte. Eur. Phys. J. C 49: 121-124, 2007
\bibitem{wiedemann} N. Borghini, U. A. Wiedemann. J. Phys. G 35: 023001, 2008
\bibitem{deadcone} Yuri L. Dokshitzer, D.E. Kharzeev. Phys. Lett. B 519:199-206,2001
\bibitem{jetmodif} N. Borghini, U. A. Wiedemann. ePrint: hep-ph/0506218

\end{thebibliography}
\end{document}